# Experimental high-dimensional entanglement certification and quantum steering with time-energy measurements

Kai-Chi Chang[1,†,*], Murat Can Sarihan[1,†], Paul Erker[2,3,†,*], Xiang Cheng[1,†], Nicky Kai Hong Li[2,3,†], Kemal Enes Akyuz[1], Andrew Mueller[4,5], Matthew D. Shaw[4], Boris Korzh[4], Maria Spiropulu[6], Marcus Huber[2,3,*], and Chee Wei Wong[1,*]

[1] Fang Lu Mesoscopic Optics and Quantum Electronics Laboratory, Department of Electrical and Computer Engineering, University of California, Los Angeles, CA 90095, USA

[2] Atominstitut, Technische Universität Wien, Stadionallee 2, 1020 Vienna, Austria

[3] Institute for Quantum Optics and Quantum Information Vienna, Austrian Academy of Sciences, Boltzmanngasse 3, 1090 Vienna, Austria

[4] Jet Propulsion Laboratory, California Institute of Technology, 4800 Oak Grove Dr., Pasadena, CA 91109, USA

[5] Applied Physics, California Institute of Technology, 1200 E California Blvd, Pasadena, CA 91125, USA

[6] Division of Physics, Mathematics and Astronomy, California Institute of Technology, Pasadena, CA 91125, USA

† These authors contributed equally to this work.

* Corresponding author email: uclakcchang@ucla.edu; paul.erker@tuwien.ac.at; marcus.huber@tuwien.ac.at; cheewei.wong@ucla.edu

**Qubit entanglement is the foundation of advanced quantum computation, non-classical information processing, sensing at the fundamental limits, and quantum communication networks. The generation of time-frequency qudit states offers increased capacities and noise robustness while keeping the number of photons constant, but poses significant challenges in the accessible measurements for certification of multi-dimensional quantum entanglement. In the time-frequency domain, for example, witnessing high-dimensional quantum steering – an intermediate non-local correlation between Bell non-locality and entanglement – has remained an open challenge. Here we demonstrate the measurement-efficient and assumption-free certification of high-dimensional entanglement with trusted measurement on each receiving node, as well as multi-dimensional semi-device-independent quantum steering using time-frequency measurement bases. First, for our 31 × 31 qudit source, we certify a lower bound of the nearly maximum entangled quantum state fidelity $\widetilde{F}(\rho,\Phi)$ of 96.2**

**± 0.2%, an entanglement-of-formation $E_{\mathrm{oF}}$ of 3.0 ± 0.1 ebits, an entanglement dimensionality $d_{\mathrm{ent}}$ of 24, and a lower bound of steering robustness $\delta(\sigma_{\mathrm{a|x}})$ of 8.9 ± 0.1. Since our qudit temporal basis is independent of the number of joint temporal measurements, our certification is achieved with the fewest number of local projective measurements to date ($d^2$ + 1; 962 for the 31 × 31 subspace), compared to the previous record of $2d^2$ (1,922 measurements). Subsequently and secondly, we are able to measure our qudit resource efficiently under dispersion conditions equivalent to 600-km of fiber transmission. After non-local dispersion cancellation, we still preserve and retrieve a 21-dimensional entanglement, with the maximum quantum state fidelity $\widetilde{F}(\rho,\Phi)$ of 93.1 ± 0.3%, an entanglement-of-formation $E_{\mathrm{oF}}$ of 2.5 ± 0.1 ebits, and a lower bound of steering robustness $\delta(\sigma_{\mathrm{a|x}})$ of 6.3 ± 0.2. Thirdly, beyond existing approaches, we efficiently measure and certify the time-frequency qudit with a new witness, via measurement bases that ensure a theoretically complete certification without perfect measurement bases. This novel practical framework, in the presence of noise and without requiring mutually-unbiased bases, shows that our qudit system has Schmidt number of at least 9 for local dimensions $d \leq 13$, matching our results in multi-dimensional quantum steering. Leveraging the intrinsic large-alphabet nature of telecom-band photons, our demonstrations enable scalable deployable entangled and steerable quantum sources, providing a pathway towards high-dimensional quantum communication networks, distributed sensing and computation.**

Quantum entanglement rose to prominence as the central phenomenon in the seminal thought experiment by Einstein, Podolsky, and Rosen (EPR) [1-3]. The EPR argument inspired Schrödinger to introduce the original concept of quantum steering, that one party can influence the wavefunction of the other party by performing suitable measurements [4]. The tangible connection from quantum entanglement to testable experiments was proposed by Bell and his now famous inequalities [5]. Once the tangible nature of the phenomenon was established, quantum entanglement emerged as a key factor in advancing numerous quantum technologies, including quantum information processing [6, 7], communication [8, 9], and computation [10, 11]. Quantum steering only recently began to receive more attention after a systematic way of understanding criteria for quantum steering was developed [12]. In the contemporary perspective, the concept of steering refers to a quantum correlation positioned between Bell non-locality and entanglement. It is alternatively referred to as a one-sided device-independent scenario [12, 13]. Notably, any quantum state that violates a Bell inequality can be utilized for steering and, while any steerable state is entangled, the reverse is not necessarily true [13, 14]. Quantum steering has since been used in fundamental quantum information processing [15-20], and asymmetric quantum communication protocols [21-24].

At present, qubit entanglement remains the predominant method employed in the majority of implementations, i.e., entanglement between two-dimensional quantum systems [6-9]. Nevertheless, recent research has unveiled the promising prospects of high-dimensional entanglement [25-58] in overcoming the limitations associated with qubit entanglement and steering. This form of entanglement and steering presents opportunities for violating local realism theories with lower detection efficiency [59-61], improved information capacities [62-72], better secure communication rates [62-64, 67-70], and higher noise resilience [67, 68, 72-74]. Attempting to harness this insight, recent experiments have achieved success in generating and certifying high-dimensional entanglement across various degrees-of-freedom (DoFs) including path [37-39, 41], orbital angular momentum (OAM) [42-48, 50], time [30, 49, 64, 68], and time-energy [25-29, 50, 60, 65, 66, 73, 74]. Nonetheless, the certification of high-dimensional entanglement and steering presents a notable hurdle. This is primarily because performing full state tomography (FST) for biphotons with a local dimension $d$ requires $(d + 1)^2d^2$ measurements with local projective bases and $(d + 1)^2$ measurements with global product bases, each having $d$ outcomes [51, 52]. Moreover, the greatest challenge comes from the fact that not every measurement is easily implementable. In time-energy entanglement, time-bins are easily accessible via accurate coincidence logics and time-taggers, but measuring superpositions of time-bins is challenging. In fact, most experiments use Franson interferometry to interfere two time-bins separated by a fixed distance, but the majority of the density matrix remains inaccessible [26, 27]. Therefore, due to the complex nature of performing measurements in high-dimensional spaces, previous experiments on certifying the entanglement dimensionality often relied on assumptions for the measured quantum state, like the conservation of OAM [44], subtraction of accidentals [56], target basis with desired correlations [42], equal contribution of diagonal elements [26], or the pure quantum state assumption of the experiments [27]. In order to fully unleash the capabilities of high-dimensional entanglement, it is essential to attain certification without making assumptions that could compromise the security and reliability of its applications. Recent advancements have made notable progress in this area, showing that experiments on two bases can enable efficient certification of qudit entanglement [54-58, 75] and quantum steering [31, 34-36, 61] without any reliance on assumptions regarding the quantum state itself. However, all these works are implemented using photonic DoFs of pixel bases [36, 54, 56, 57, 61, 75], path [58], or OAM [31, 34, 35, 55]. So far, different approaches have been tested in time-frequency DoF, such as Franson interferometers to interfere neighboring time-bins [26, 27, 73], or electro-optic phase modulators for interfering with specific frequency-bins [40, 53]; however, these methods have limitations on scalability in terms of number of accessible measurements. Furthermore, the certification of high-dimensional quantum steering in time-energy DoF remained a scientific desideratum. Moreover, the existing certification methods for high-dimensional entanglement with local measurement bases typically require specific measurement bases [55, 57].

Here we demonstrate the measurement-efficient and assumption-free certification of high-dimensional entanglement with trusted measurement devices, as well as the first multi-dimensional quantum steering in a semi-device-independent manner using time-frequency bases. We present a general approach to prepare and manipulate time-frequency bases of two-photons independently. By utilizing large-alphabet temporal encoding and fiber-optics telecommunications components combined with our low-jitter and high-efficiency single-photon detectors, we efficiently generate a $31 \times 31$ dimensional time-frequency basis to certify high-dimensional entanglement with optimal witness and quantum steering inequalities. Our time-frequency bases approach allows the fewest number of local projective measurements at $d^2+1$ (962 measurements for a $31 \times 31$ dimensional subspace), due to the $d$-outcome measurements in the temporal domain. This $d$-outcome measurement approach is independent of the number of joint temporal measurements, enabling the scaling up in dimensionality for a single measurement setting. Subsequently we are able to certify a lower bound of the fidelity with a nearly maximally entangled state $\tilde{F}(\rho,\Phi)$ at 96.2 ± 0.2%, an entanglement-of-formation $E_{oF}$ at 3.0 ± 0.1 ebits, and an entanglement dimensionality $d_{\text{ent}}$ of 24 respectively. We extract a lower bound steering robustness $\delta(\sigma_{\mathrm{a|x}})$ of 8.9 ± 0.1 with a certified Schmidt number $n$ equal to 9, thereby demonstrating 9-dimensional quantum steering. Subsequently and secondly, we demonstrate efficiently the preservation of qudit nature through non-local dispersion cancellation, a concept in time-energy entanglement for clock synchronization [77, 78] and secure communication. After long-distance entanglement transport through a ±10,000 ps/nm dispersion emulator (compensator) with 600-km equivalent distances, non-local dispersion cancellation still affords the maximum quantum state fidelity $\tilde{F}(\rho,\Phi)$ of 93.1

± 0.3%, entanglement-of-formation $E_{oF}$ at 2.5 ± 0.1 ebits, and entanglement dimensionality $d_{\text{ent}}$ of 21 respectively. We achieve a lower bound of steering robustness $\delta(\sigma_{\mathrm{a|x}})$ of 6.3 ± 0.2, corresponding to a Schmidt number $n$ of 7 certification and demonstrating a 7-dimensional quantum steering. Thirdly, we experimentally demonstrate a novel certification framework [83] across a broad range of local dimensions $d$. For instance, with measurement bases of $d \leq 13$, we witnessed the state Schmidt number to be at least 9 while their infidelity with the maximally entangled state, $1 - \mathcal{F}(\rho, \Phi_d^+)$, is at most 33.68%. The upper bounds of the allowable experimental noise ratio and the dimension-rescaled bases are also measured, matching our theoretical models. Our work provides a step towards achieving large-scale quantum information processing and noise-tolerant high-capacity quantum networks in a scalable and fiber-optic telecommunication platform.

## Results

### Generation of two bases using discretized time-frequency subspaces

To generate high-dimensional entanglement and quantum steering in the photonic time-frequency DoF, we use spontaneous parametric down-conversion (SPDC), a second-order light-matter nonlinearity that mediates the annihilation of one pump photon, simultaneously generating two daughter photons, typically referred to as signal and idler photons [6, 7, 52]. SPDC gives rise to photon-pairs that preserve the energy, momentum, and polarization of the incident optical field, creating two-photons in a continuum of time and frequency modes. Consequently, this results in strong quantum correlations observed in the joint temporal intensity (JTI) and joint spectral intensity (JSI), as illustrated in Figure 1a. By utilizing the arrival-time large-alphabet encoding and telecom-band frequency filtering techniques, we can discretize the time and frequency modes of SPDC and generate two bases as shown in Figure 1a. For discretized JTI, we use high-dimensional temporal encoding with our correlated SPDC photon-pairs as noted in Figure 1b. The adjacent light blue slots in the diagram represent local timing jitter errors. To effectively control the JTI, two key parameters come into play: the bin-width $\tau$ and the number of bins $N$. Careful consideration should be given to select these parameters to fully utilize the available photon detection resource and optimize the JTI measurements. Purple slots indicate an example that there are no coincidence photons that can be registered. For the generation of discretized JSI, we utilize commercial telecom-band frequency filtering to individually select frequency-correlated photon-pairs. In our scheme, we generate an entangled and steerable qudit state with our SPDC source, where the signal and idler photons are distributed to each party, Alice and Bob, respectively. Hence the discretized JTI can be fully controlled in the temporal domain, while the discretized JSI can be independently controlled in the frequency domain. Generating two bases with the discretized JTI and JSI allows us to certify high-dimensional entanglement with a fidelity lower bound [55, 57], entanglement-of-formation [26, 27, 55, 57], achieve the expectation value via a new witness [83], and to certify high-dimensional quantum steering using lower bound of steering robustness and certified Schmidt number [35, 36], as presented in Figure 1c. Figure 1d summarizes the canonical number of measurements required to certify high-dimensional entanglement via optimal full state tomography (FST), optimal fidelity $F(\rho,\Phi^{+})$, fidelity bounds $\tilde{F}$ $(\rho,\Phi^{+})$ [55], and our efficient $d^2$+1 time-frequency bases measurements noted in this work. The efficient 1+$d^2$ certification arises because the JTI measurements only require a single (1) measurement setting (having $d$-outcomes from the time-bins) and the JSI measurements are still projection measurements requiring $d \times d$ measurement settings ($d^2$) from two swept local spectral filters.

With the efficient measurement approach, we next scaled our discretized JTI to 256 × 256 dimensions as noted in Figures 2a and 2b. Both Alice and Bob use their fiber beamsplitters with 50:50 ratios for two-photon temporal correlation measurements ($T_A$ and $T_B$) and spectral correlation measurements ($F_A$ and $F_B$), each detected by two low-jitter and two high-efficiency superconducting nanowire single-photon detectors (SNSPDs). The Methods section describes further details on the experimental setup, as well as information on the low-jitter [81, 82] and high-efficiency SNSPDs. The number of bins $N$ is 256 and the bin-width $\tau$ is fixed at 250 ps, carefully chosen to fully cover the entire two-photon correlation peak (Figure 2b right-middle inset) which bounded by the detector and electronic jitter of our coincidence counting module. With the optimized bin-width $\tau$, the cross-talk in the off-diagonal elements is small compared to the diagonal elements. Effects of the bin-width $\tau$ and number of bins $N$ are further illustrated in Supplementary Information Figure S1 and S2. Figure 2b embedded inset also shows non-optimal 7-dimensional JTI when the bin-width $\tau$ is chosen to be 31.6 ps for comparison purpose. We note that discretized JTI measurement time is fixed at 3 seconds in a single setting, regardless of the subspace dimension, with data post-processing to generate the JTIs. This approach furthers allows the dimensionality scaling with SNSPDs with lower timing jitter (allowing even smaller bin-widths $\tau$) while increasing the number of bins $N$, supporting larger-alphabet temporal encoding and higher secure key capacity [62, 63, 67, 68]. Here we note that the different dimensions of JTI all come from a single measurement setting, and with post-processing of the data enables the generation of these JTIs. In contrast, for all the JSI measurements, we use the traditional projection measurements that scale as $2d^2$. Hence, in this work, our proposed time-frequency bases provide us with a $d^2$+1 scaling in terms of the local measurement settings. Figure 2c shows the measured 31 × 31 dimensional discretized JSI using a pair of tunable frequency filters for protection measurements between Alice and Bob. We align the central wavelength of a pair of frequency filters to the center of our SPDC photons and sweep the frequency symmetrically with respect to the center wavelength to register coincidence counts from two highly-efficient SNSPDs. The coincidence counts fall-off is due to the SPDC phase-matching bandwidth ≈ 250 GHz. Inset of Figure 2c is the measured discretized 7-dimensional JSI by using a pair of 5.9 GHz FWHM tunable frequency filters ($F_A$ and $F_B$).

We perform the verification of mutually-unbiased bases using the cross-basis measurement of our time-frequency bases. Two $d$-dimensional bases, indexed by $m$ and $n$, are said to be mutually-unbiased, if their constituting elements, indexed by $i$ and $j$, satisfy the following relation [43, 64]:

$$\left|\langle\psi_{m,i}|\psi_{n,j}\rangle\right|^2 = \begin{cases} 1/d \;\; for\; m \neq n \\ \;\delta_{ij} \;\; for\; m = n \end{cases} \quad (1)$$

for all $i$ and $j$. This implies that when experiments are conducted in two mutually-unbiased bases, the outcomes obtained in one basis provide minimal information of the corresponding results for the other basis. As we do not make assumptions about our measurement bases, we additionally test the unbiasedness by measuring cross-detection probabilities. Using our time-frequency and frequency-time measurements (see Figure S3 in Supplementary Information for more details), for a 7-dimensional subspace, we extract a joint cross-detection probabilities of 0.14812 for $m \neq n$, which is close to the value of 1/7 (0.14285) for an ideal 7-dimensional mutually-unbiased bases. Moreover, we compute the average deviation from a joint cross-detection probability of an ideal 7-dimensional mutually-unbiased bases measurement to be $0.00202 \pm 0.00178$.

**Witness of high-dimensional entanglement and quantum steering with time-frequency bases**

To witness high-dimensional entanglement, we use a fidelity bound [55] and an entanglement-of-formation bound [57], where both methods can establish the bound via two measurement bases. To measure the high-dimensional quantum steering, we utilize our proposed time-frequency bases and a recent approach that determines a lower bound of steering robustness [36], which also yields the certified Schmidt number of the state at hand. We also certify the Schmidt number of our states with the newly established Schmidt number witness [83] that utilizes correlations in measurement bases, providing a theoretically complete Schmidt number certification even if one cannot control the measurement bases perfectly. The spirit of fidelity bound certification method is that one needs measurements in at least two distinct bases to certify high-dimensional entanglement [55]. Considering a given quantum state $\rho$ characterized by a Schmidt number not exceeding $k$, an inequality formula can be derived:

$$\tilde{F}(\rho,\Phi) \leq F(\rho,\Phi) \leq B_k\,(\Phi) \tag{2}$$

where $\tilde{F}(\rho,\Phi)$ is the quantum state fidelity's lower bound, $F(\rho,\Phi)$ is the obtained quantum state fidelity, and $B_k\,(\Phi)$ is the witness cutoffs for the quantum state of Schmidt number $k$. Using time-frequency bases, we can apply this approach to establish such fidelity $\tilde{F}(\rho,\Phi)$'s lower bound, and the certified entanglement dimensionality $d_{\text{ent}}$, given by the maximum Schmidt number $k$ for a specific quantum state $\rho$. Similarly, by using measurements in two distinct bases, it is proven that one can bound the entanglement-of-formation $E_{\text{oF}}$ of a quantum system [57] by:

$$E_{\text{oF}} \geq -\log_2\left(\max_{i,j}(|\langle i|\tilde{j}\rangle|^2)\right) - H(A_1|B_1) - H(A_2|B_2) \tag{3}$$

where $H(A_1|B_1)$, $H(A_2|B_2)$ are the conditional Shannon entropy for outcomes in the first and second bases respectively, and $\max_{i,j}(|\langle i|\tilde{j}\rangle|^2)$ is the maximal overlap of elements of the two bases used (which would be $1/d$ in case of ideal mutually-unbiased bases). An evaluation of the measurement outcomes of our setup leads to $\max_{i,j}(|\langle i|\tilde{j}\rangle|^2)$ of 0.14812 (more details in Supplementary Information). The entanglement-of-formation $E_{\text{oF}}$ is the minimal number of maximally entangled two-qubit states needed to create quantum state $\rho$ via classical communications and local operations [26, 54]. A pair of two-dimensional quantum systems can contain at most 1 ebit (corresponding to a maximally non-separable qubit system), while high-dimensional systems can contain up to $\log_2(d)$ as an approach towards high-dimensional certification. Here we further certify in a stricter form – quantum steering – to witness the high-dimensionality of our system. Recently a method has been proposed and conducted in a genuine high-dimensional quantum steering using pixel basis [47]. We employ this approach to validate the quantum steering with higher-dimensionality by our time-frequency bases through following violation criteria:

$$n \geq \left(\frac{1+\mathrm{SR}(\sigma_{\mathrm{a|x}})}{1-\mathrm{SR}(\sigma_{\mathrm{a|x}})}\right)^2 \equiv \delta(\sigma_{\mathrm{a|x}}) \tag{4}$$

where $n$ is the certified Schmidt number via steering, SR is the steering robustness, $\delta(\sigma_{a|x})$ is the SR's lower bound, and $\sigma_{a|x}$ is the assemblage. Equation (4) indicates the feasibility of extracting a lower bound of certified Schmidt number $n$ from the maximum integer value of $\delta(\sigma_{a|x})$ to witness multi-dimensional quantum steering. This can be achieved by utilizing the coincidence counts obtained from measurements conducted in two bases.

Figure 2d shows the high-dimensional entanglement witness using the outcomes of our 31 × 31 dimensional discretized JTI and JSI measurements. Note that we use the 31 × 31 matrix for certification not the full 256 × 256, since our complementary JSI measurements are bounded by 5.9 GHz tunable frequency filters bandwidth and the ≈ 250 GHz SPDC source phase-matching bandwidth. For a local dimension $d$ of 3, we witness a maximum quantum state fidelity $\tilde{F}(\rho,\Phi)$ up to 96.2 ± 0.2%, and an entanglement-of-formation $E_{oF}$ of 1.3 ± 0.1 ebits for a entangled dimensionality $d_{ent}$ of 3. For a local dimensional subspace $d$ at 23 × 23, we extract a lower bound quantum state fidelity $\tilde{F}(\rho,\Phi)$ of 82.1 ± 0.3% (for example, for local dimension $d$ = 18, our threshold $B_k$ ($\Phi$) is 78.3% and thus here we have a entangled dimensionality $d_{ent}$ of 19), and an entanglement-of-formation $E_{oF}$ of 3.0 ± 0.1 ebits. By expanding the dimensions in our time-frequency bases to 31 × 31, in Figure 2d we successfully witness a lower bound fidelity of $\tilde{F}(\rho,\Phi)$ of 77.0 ± 0.2%, with an entanglement-of-formation $E_{oF}$ of 3.0 ± 0.1 ebits. Figure 2e presents the one-to-one relationship between the local dimensionality $d$, and the certified entanglement dimension $d_{ent}$, where we observe a witness of a 24-dimensional entangled dimensionality $d_{ent}$ when local dimensionality $d$ is 31. The uncertainty in fidelity is determined from each measurement data set, assuming Poisson statistics. As a proof-of-concept demonstration, our quantum state fidelity $\tilde{F}(\rho,\Phi)$, entanglement-of-formation $E_{oF}$, and entangled dimensions $d_{ent}$ are higher than recent studies [27, 29], and comparable with the current record of dimension witnesses without accidental subtraction [26, 57].

We further utilize our 31 × 31 dimensional time-frequency bases to certify high-dimensional quantum steering. In Figure 2f, we present this result. For a local dimension $d$ of 3, we witness a $\delta(\sigma_{a|x})$ of 2.7 ± 0.04, and hence the certified Schmidt number $n$ is 3 [from maximum integer value of the quantity $\delta(\sigma_{a|x})$, as given by Equation (4)], demonstrating a 3-dimensional steerable quantum state. Similar to the high-dimensional entanglement witness, we measure quantum correlations in two bases for time-frequency subspaces of dimension up to $d$ = 31. For a local dimension $d$ of 23, we achieve a steering robustness lower bound $\delta(\sigma_{a|x})$ up to 8.9 ± 0.1 to certify a 9-dimensional quantum steering. Both the high-dimensional entanglement and quantum steering witnesses are derived from the consistent raw data of Figure 2b top-left inset and 2c, and our results are in-line with the fact that quantum steering is a stricter correlation than entanglement [11, 12]. Given the same two bases' data, the maximum dimension we can certify will generally be lower for high-dimensional quantum steering compared to high-dimensional entanglement. For all the presented results here, accidental coincidence counts subtraction is not applied. In Figure S4 in Supplementary Information, we further analyze the individual Schmidt eigenvalues from JTI and JSI matrices in Figure 2b top-left inset and 2c, using the Schmidt mode decomposition method [27]. We find out that the corresponding Schmidt eigenvalues for JTI matrices decrease slower than those of the JSI matrices, with better uniformity, and hence more favorable towards higher dimensionality. Moreover, we also point out a possible future improvement on the JSI measurements towards higher-dimensional states as shown in Figure S5 and in Table I of the Supplementary Information. In Figure 2b and 2c, the measured coincidence counting duration is 3 seconds, and no background counts subtraction is applied.

**Time-frequency high-dimensional entanglement and quantum steering preservation after ± 10,000 ps/nm non-local dispersion cancellation**

The transmission of fragile quantum correlations with a noisy channel represents a fundamental challenge in the realm of quantum communication [8, 9], and quantum imaging [79, 80], with dispersion affecting the signal and idler photon correlation after long distances. Here we proceed to demonstrate the time-frequency high-dimensional entanglement and quantum steering preserving after non-local dispersion cancellation. Figure 3a shows the measurement setup with ±10,000 ps/nm dispersion emulator and compensator (Proximion) using chirped fiber Bragg gratings, with equivalent net dispersion values of ≈ 600-km of standard single-mode fibers with ≈ 3 dB loss. In Figure 3b, we show the two-photon coincidence counts from 107 × 107 dimensional discretized JTI after ± 10,000 ps/nm nonlocal dispersion cancelation. The bin-width $\tau$ is chosen at 600 ps to fully cover the entire two-photon correlation peak (Figure 3b top-middle inset), and the number of bins $N$ is 107, to be consistent with the time frame size that we use in Figure 2b. After non-local dispersion cancellation, the FWHM of temporal correlation peak is observed to be ≈ 128.7 ps, and the asymmetry profile comes from the imperfect non-local dispersion cancellation. For a 107-dimensional JTI, we observe that the coincidence counts of temporal correlated photons are consistent with the level of coincidence counts for a 31-dimensional JTI shown in Figure 3b top-left inset. This clearly demonstrates the advantage of our scheme. Comparing Figure 3b to 2b, we observe that after ±10,000 ps/nm non-local dispersion cancellation, the JTI matrix becomes more noisy, which is mainly due to the insertion loss of dispersion modules, and the imperfect non-local dispersion cancellation. We emphasized again that, in Figure 2a and 3b, all the JTI measurements are performed in 3 seconds with a single measurement setting (*d*-outcome measurement). Due to the large-alphabet temporal encoding scheme, the number of measurements and the measurement time for discretized JTI is fixed as long as there are sufficient registered coincidence counts. In our case, the typical coincidence-to-single counts ratio is about 1:9 for JTI measurements, regardless of the matrix dimension. For comparison purpose, in Figure 3b bottom-right inset, we illustrate a typical noisy 31-dimensional JTI when +10,000 ps/nm dispersion is not compensated.

After such large non-local dispersion cancellation, we analyze the high-dimensional entanglement witnesses using fidelity bounds and entanglement-of-formation bounds from outcomes of the measured time-frequency bases. This is detailed in the Figure S6 of Supplementary section IV. For a local dimension $d$ of 19, we obtain a transported entangled dimensionality $d_{ent}$ of 15 with a state fidelity $\tilde{F}(\rho,\Phi)$ of 77.0 ± 0.4%, and an entanglement-of-formation $E_{oF}$ of 2.5 ± 0.1 ebits. Through additional enhancements in the dimensionalities of our time-frequency bases, we have managed to transmit a lower bound quantum state fidelity $\tilde{F}(\rho,\Phi)$ of 65.9 ± 0.3% and an entanglement-of-formation $E_{oF}$ of 2.3 ± 0.1 ebits, with uncertainty from the Poissonian statistics. Figure 3c shows the corresponding dispersion-cancelled high-dimensional quantum steering results. For a local dimension $d$ of 3, we witness a steering robustness lower bound $\delta(\sigma_{a|x})$ of 2.4 ± 0.1 and hence the certified Schmidt number $n$ is 3 for a transported 3-dimensional steerable state. For a local dimension $d$ of 19, we achieve a steering robustness lower bound $\delta(\sigma_{a|x})$ up to 6.3 ± 0.2 to certify a 7-dimensional quantum steering after non-local dispersion cancellation. The decrease of high-dimensional entanglement and steering witness is mainly due to the bandwidth limitation of our SPDC source and the increase noise photons for discretized JTI after large non-local dispersion cancellation.

**High-dimensional entanglement certification via new witness for time-frequency bases**

Here we present experimental data for a new witness with time-frequency bases, first proposed in [83]. We cross-check the verified Schmidt numbers $n$ of our states with this proposed witness.

By measuring the state of interest $\rho$ in $m$ coordinated local orthonormal bases ($m$ = 2 in this experiment for time and frequency bases), we obtain the witness expectation value $S_d^{(m)}(\rho)$ (defined in Methods) which satisfies the inequality:

$$S_d^{(m)}(\rho) \leq \frac{k(m-\mathcal{T}(\mathcal{C}))}{d} + \mathcal{T}(\mathcal{C}) \equiv B_k \tag{5}$$

if $\rho$ has Schmidt number at most $k$, where the quantity $\mathcal{T}(\mathcal{C})$ depends only on all the absolute-value squared of the measurement bases overlaps (see Methods for complete descriptions). The violation of Inequality (5) certifies the state $\rho_{AB}$ to have Schmidt number $n$ of at least $k+1$. Furthermore, the two quantities $S_d^{(m)}(\rho)$ and $\mathcal{T}(\mathcal{C})$ also allow us to lower bound the entanglement fidelity $\mathcal{F}(\rho, \Phi_d^+)$ which is the maximum fidelity between $\rho$ and all (local-unitarily equivalent) maximally entangled states (details are presented in Methods).

Figure 4a presents our certification results of the high-dimensional Schmidt number witness with time-frequency bases. The certified Schmidt numbers $n$ is mostly higher than (with a few local dimensions equal to) the ones certified using the quantum steering robustness [36] as in Figure 2f, demonstrating a unique advantage of our witness. Due to the generality of the witness derived in [83], it is sensitive to noise which bounds the certification for nontrivial Schmidt numbers $n$ of larger dimensions $d$ (for example, greater than 17). Thus, in Figure 4b we compare the witness expectation value $S\rho_{exp}^d \coloneqq S_d^{(m)}(\rho_{exp})$ of the state from our experiment $\rho_{exp}$ with $S\rho_{iso}^d \coloneqq S_d^{(m)}(\rho_{iso})$ of the isotropic state $\rho_{iso} = (1-p)|\Phi_d^+\rangle\langle\Phi_d^+| + \frac{p}{d^2}\mathbb{1}_{d^2}$ for various white-noise ratios $p$, where both states are measured in our time-frequency bases ($m = 2$). We plot these witness-expectation values and the upper bounds for any state to have Schmidt number, $B_k$, for local dimension $d$ = 7, where the certified Schmidt number $n$ is the same as the local dimension. The horizontal position of the experimental datapoint informs the white-noise ratio of an isotropic state that corresponds to the same witness expectation value of the state from our experiments, suggesting that our experimental state is as entangled as a low white-noise isotropic state from the perspective of the Schmidt number witness. For all the other experimental witness results, we presented them in Figure 4c. In Figure 4c, we show the complete and explicit values for the witness expectation value $S_d^{(m)}(\rho)$, the witness upper bound $B_k$ corresponding to the largest $k$ such that $S_d^{(m)}(\rho) > B_k$, and the highest certified Schmidt number in each dimension (illustrated in Figure 4a and 4b). This is computed using the measurement results in Figure 2b top-left inset and 2c, and the cross-basis measurements in Figure S7 of Supplementary Information. We observe that as the dimension of our time-frequency bases increases, the experimental system has higher noise and hence the witness expectation value $S_d^{(m)}(\rho)$ decreases. Even with higher noise, this certification framework is still able to preserve and certify a 9-dimensional entanglement using our time-frequency bases measurements.

Furthering from the qudit certification, we implement a new high-dimensional entanglement witness with our time-frequency bases. Figure 4d presents the upper bounds of the white-noise ratio $p_{c,m=2}^{k}$ ( $m = 2$ ) for experimentally certified Schmidt number $n$ versus $\varepsilon_{min}$ (which quantifies the deviation of the measurement bases from mutually-unbiased bases; see Methods for more details) for an example local dimensionality $d$ of 17. We superimpose the measurement data (stars) with the theoretical predictions for dimensions $d$ from 3 to 17. Note that some of the data point will be in a same curve since they have same certified Schmidt number $n$ (for example, the certified Schmidt number 9 is shared between $d = 11$ and 13), but with different white-noise ratio $p_{c,m=2}^{k}$. The white-noise ratio $p_{c,m=2}^{k}$ of the measurement data is obtained from the experimental witness expectation value $S_d^{(m)}(\rho)$ at different dimensions. The noise tolerance of this witness is higher for smaller $k$ but reduces as $\varepsilon_{min}$ increases. Moreover, Figure 4e plots the computed suprema of the dimension-rescaled bases bias $d\varepsilon_{min}$versus the dimensionality $d$ using our time-frequency bases. From Figure 4d we extract the $\varepsilon_{min}$ for each dimensionality $d$, and subsequently, by using the experimental certified Schmidt numbers in Figures 4a and 4c, we compute the best fit between theoretical curves and our measurement dataset. The measurement data (stars) are illustrated together with the theory for dimensions $d$ from 9 to 17. The white-noise ratio $p$ we used for Figure 4e is 0.058, close to the white-noise ratio $p$ of 0.05 in local dimension $d = 7$, where the certified Schmidt number is the same as the local dimension, as shown in Figures 4b and 4c. This is due to the difference between our experimental time-frequency quantum state and the isotropic state. We also provide the supporting measurements in Figure S8, and the certification results for experimental data after ± 10,000 ps/nm non-local dispersion cancellation in Figure S9 of the Supplementary Information.

**Conclusion**

We summarize our results in comparison with other reported works. First, we can access all measurements data that scale as ($d^2 + 1$) for certifying higher dimensional quantum states without any physical assumptions on quantum state itself, which is more general compared to previous works [26, 27, 42, 44]. Second, the time-frequency quantum steering dimensions that we certify in this work have not been observed prior. Our time-frequency measurement bases have shown maximum quantum state fidelity $\tilde{F}(\rho,\Phi)$ of 96.2 ± 0.2%, an entanglement-of-formation $E_{\mathrm{oF}}$ of 3.0 ± 0.1 ebits, an entanglement dimensionality $d_{\mathrm{ent}}$ of 24 at the qudit source; and 21-dimensional entanglement, with the maximum quantum state fidelity $\tilde{F}(\rho,\Phi)$ of 93.1 ± 0.3%, an entanglement-of-formation $E_{\mathrm{oF}}$ of 2.5 ± 0.1 ebits after non-local dispersion cancellation of 600-km dispersion-equivalent distance. Based on these results, we also present records of witnessed high-dimensional time-energy entanglement compared to recent works [27, 29]. Third, we utilized the recently developed witness [83] to certify our high-dimensional entanglement experimental systems, for the first time, to the best of our knowledge. Our qudit system has certified Schmidt number $n$ of 9, matching our certification results in multi-dimensional quantum steering and is quantified in the presence of noise. We note that this framework has the capability of establishing certification bounds even when the measurement bases are not exactly mutually-unbiased, extending the capabilities from prior experiments. Fourth, compared with prior studies based on spatial mode DoFs, our discretized JTI and JSI can be generated and independently controlled in a single spatial mode that is directly applicable to current telecommunication fiber infrastructure, helpful for our scheme to be implemented in large-scale quantum platforms and in high-rate noise-robust quantum networks.

**Data availability and Materials**

The datasets generated and analyzed during this study are available from the corresponding authors upon reasonable request.

**Funding and Acknowledgement**

The authors acknowledge discussions with Natalia Herrera Valencia, Sébastien Designolle, Patrick Hayden, and discussions on the superconducting nanowire single-photon detectors with Vikas Anant. This study is supported by the Army Research Office Multidisciplinary University Research Initiative (W911NF-21-2-0214), National Science Foundation under award numbers 1741707 (EFRI ACQUIRE), 1919355, 1936375 (QII-TAQS), and 2137984 (QuIC-TAQS). Part of this research was performed at the Jet Propulsion Laboratory, California Institute of Technology, under contract with NASA. P.E., N.K.H.L. and M.H. acknowledge support from the European Research Council (Consolidator grant "Co-coquest" 101043705), the European flagship on quantum technologies ("AS- PECTS" consortium 101080167), Fixie (FQXi- IAF19-03-S2, within the project "Fueling quantum field machines with information") and from the European Commission (grant 'Hyperspace' 101070168). P.E., N.K.H.L. and M.H. further acknowledge support from the Austrian Federal Ministry of Education, Science and Research via the Austrian Research Promotion Agency (FFG) through the flagship project FO999897481 (HPQC) and through Quantum Austria projects 914033 (QUICHE) and 914030 (MUSIQ) funded by the European Union – NextGenerationEU. N.K.H.L. acknowledges financial support from the Austrian Science Fund (FWF) through the stand-alone project P 36478-N funded by the European Union – NextGenerationEU.

**Author contributions**

K.-C.C. developed the idea and design the experiments. K.-C.C., X.C., and M.C.S. conducted the measurements. K.-C.C., M.C.S., and N.K.H.L. contributed to the data analysis. P.E., N.K.H.L., M.H., and K.-C.C. contributed to theoretical calculations. A.M., M.S., M.D.S., and B.K. contributed the low-jitter SNSPD detectors. M.H., P.E., X.C., M.C.S. K.E.A, and C.W.W. supported and discussed the studies. K.-C.C., P.E., N.K.H.L., M.H., and C.W.W. prepared the manuscript. All authors contributed to the discussion and revision of the manuscript.

**Competing interests**

The authors declare no competing interests.

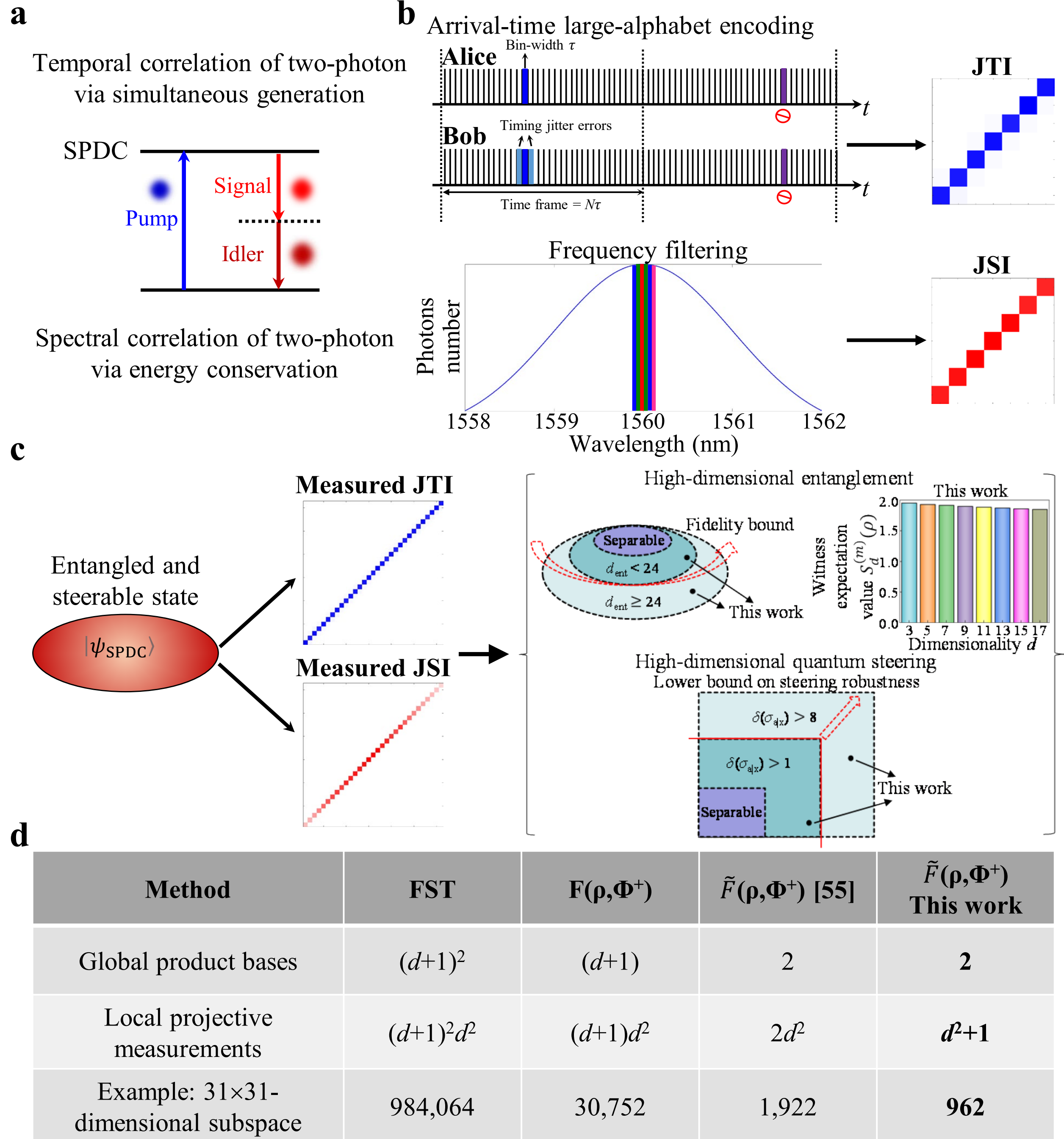

| Method | FST | $F(\rho,\Phi^+)$ | $\tilde{F}(\rho,\Phi^+)$ [55] | $\tilde{F}(\rho,\Phi^+)$ This work |
|---|---|---|---|---|
| Global product bases | $(d+1)^2$ | $(d+1)$ | 2 | **2** |
| Local projective measurements | $(d+1)^2d^2$ | $(d+1)d^2$ | $2d^2$ | $\mathbf{d^2+1}$ |
| Example: 31×31-dimensional subspace | 984,064 | 30,752 | 1,922 | **962** |

**Figure 1 | Principle of high-dimensional entanglement and quantum steering certification using efficient time-frequency bases. a,** Spontaneous parametric down-conversion (SPDC) is a second-order nonlinear process that has the annihilation of one photon from a pump field and the simultaneous generation of two daughter twin photons, typically referred to as signal and idler photons. The simultaneous two-photon correlated generation dictates that if one photon is detected at a time step, the other photon must be at the same time step, leading to a strong correlation in the joint temporal intensity (JTI). This process also preserves the energy; the sum of the signal and idler photon frequencies is constant and hence energy conservation yields a strong correlation in the joint spectral intensity (JSI). **b,** The more detailed discretized JTI and JSI generation from

continuous time-frequency modes in SPDC. For discretized JTI, we use high-dimensional temporal encoding with our correlated photon-pairs. Timing jitter errors are represented by light blue slots and the two key parameters to control JTI are the bin-width $\tau$ and the number of bins $N$, which should be chosen to fully utilize the available photon resource. Purple slots indicate that there are no coincidence photons which can be registered. For discretized JSI, we utilize commercial telecom-band frequency filtering to select the frequency correlated photon-pairs. **c,** In this work, via discretizing JTI and JSI measurements, we can certify both high-dimensional entanglement with fidelity lower bound, and high-dimensional Einstein-Podolsky-Rosen (EPR) steering using lower bound of steering robustness. In addition, we show a new experimental witness expectation value $S_d^{(m)}(\rho)$ for various dimensionality $d$, the foundation to obtain bounds on the certified Schmidt number and upper bound of entanglement fidelity for time-frequency bases. **d,** Our time-frequency joint approach allows the most efficient certification of high-dimensional entanglement to date, even when compared with optimal full state tomography (FST), optimal fidelity measurement $F(\rho,\Phi^+)$, and fidelity bounds measurements $\tilde{F}(\rho,\Phi^+)$.

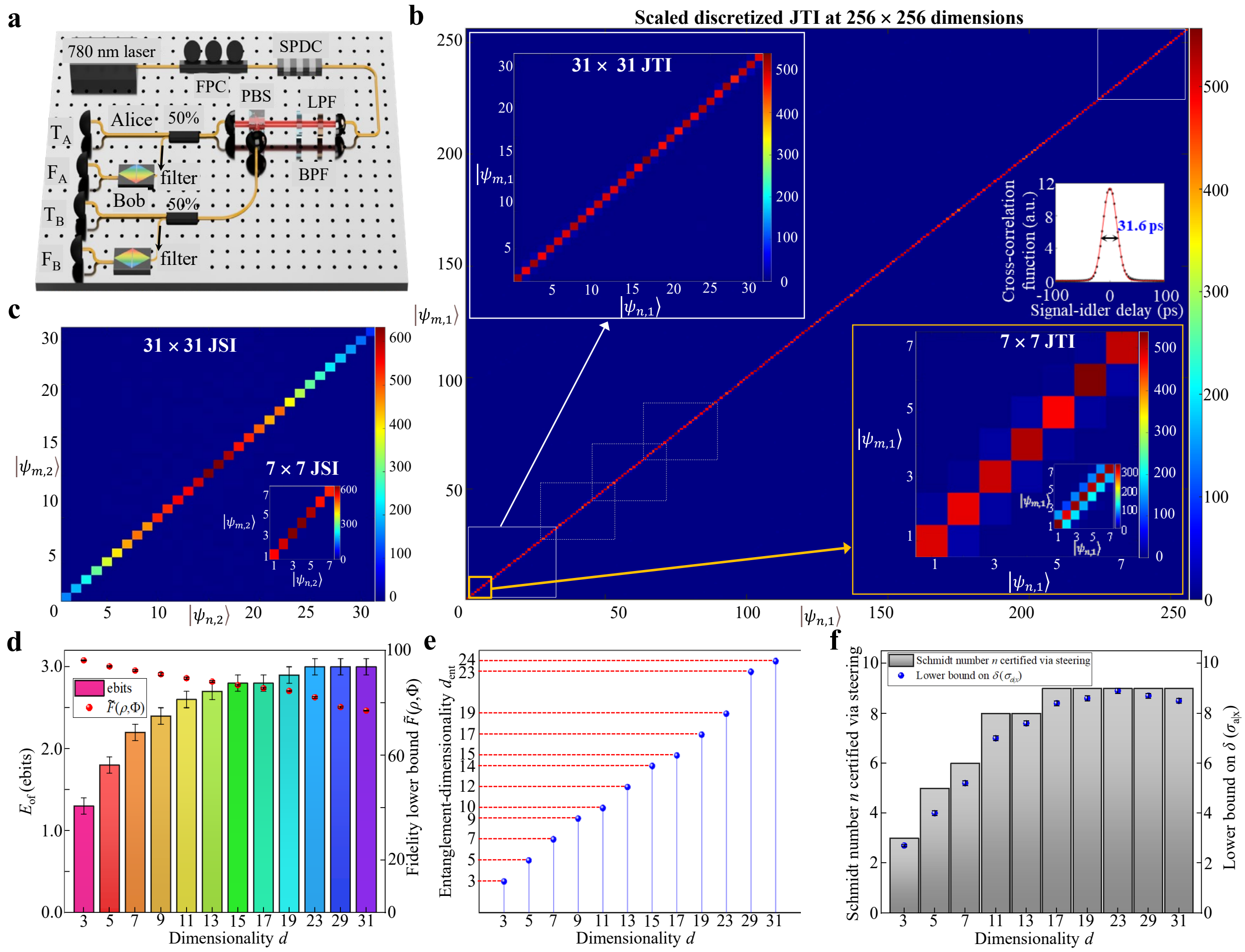

**Figure 2 | Efficient certification of high-dimensional entanglement and quantum steering using time-frequency measurements. a,** Measurement schematic. FPC: fiber polarization controller; LPF: long-pass filter; BPF: band-pass filter; and PBS: polarization beam-splitter. After separating signal and idler photons, both Alice and Bob use their fiber beam splitters with 50:50 ratio for two-photon temporal ($T_A$ and $T_B$) and spectral ($F_A$ and $F_B$) correlation measurements, detected by two low-jitter and two high-efficiency superconducting nanowire single-photon detectors (SNSPDs). **b,** The resulting measured 256 × 256 dimensional discretized JTI. With optimized bin-width $\tau$ of 250 ps in our measurements, the cross-talk in the off-diagonal elements is small compared to the diagonal elements. The 250 ps bin-width is chosen to fully cover the entire two-photon correlation peak, as shown in the right-middle inset. The number of bins $N$ at 256 and can be scaled up with larger time frame size. Top-left inset: a 31 × 31-dimensional JTI subset which shows the consistent coincidence counts with the larger 256 × 256-dimensional JTI. Bottom-right inset: a corresponding smaller 7 × 7-dimensional JTI subset at 250 ps bin-width. For comparison purposes, when the bin-width is non-optimal at 31.6 ps, the off-diagonal counts are significant as shown in the embedded inset. Note that we use the 31 × 31 matrix for certification (not the full 256 × 256), since our complementary JSI measurements are bounded by 5.9 GHz tunable frequency filters bandwidth and the ≈ 250 GHz SPDC source phase-matching bandwidth. **c,** A measured 31 × 31-dimensional discretized JSI using a pair of 5.9 GHz FWHM tunable frequency filters ($F_A$ and $F_B$) to perform projection measurements between Alice and Bob. The

coincidence counts fall-off is due to the phase-matching bandwidth in our SPDC source, about 250 GHz. Inset is the measured discretized 7 × 7-dimensional JSI. **d,** With the discretized 31 × 31-dimensional JTI and JSI at hand, we perform high-dimensional entanglement witness using fidelity and entanglement-of-formation bounds from measurements of two bases. The maximum lower bound quantum state fidelity $\tilde{F}(\rho,\Phi)$ and entanglement-of-formation $E_{oF}$, obtained are 96.2 ± 0.2%, and 3.0 ± 0.1 ebits respectively. The uncertainties are calculated each measurement data set, assuming Poissonian statistics. **e,** The relation between local dimension $d$ and the actual certified entanglement dimensionality $d_{ent}$ is provided here. The maximum certified entanglement dimensionality $d_{ent}$ is 24 from our measurements. **f,** High-dimensional quantum steering certification via the 31 × 31 time-frequency bases. In a local dimension $d$ of 19, we extract the steering robustness (SR) lower bound $\delta(\sigma_{a|x})$ of 8.9 ± 0.1. The certified Schmidt number $n$ is 9 which demonstrates a 9-dimensional quantum steering, without requiring assumptions on the quantum state. For all the results present here, no subtraction of background or accidental counts is performed. The coincidence count measurement duration is 3 seconds.

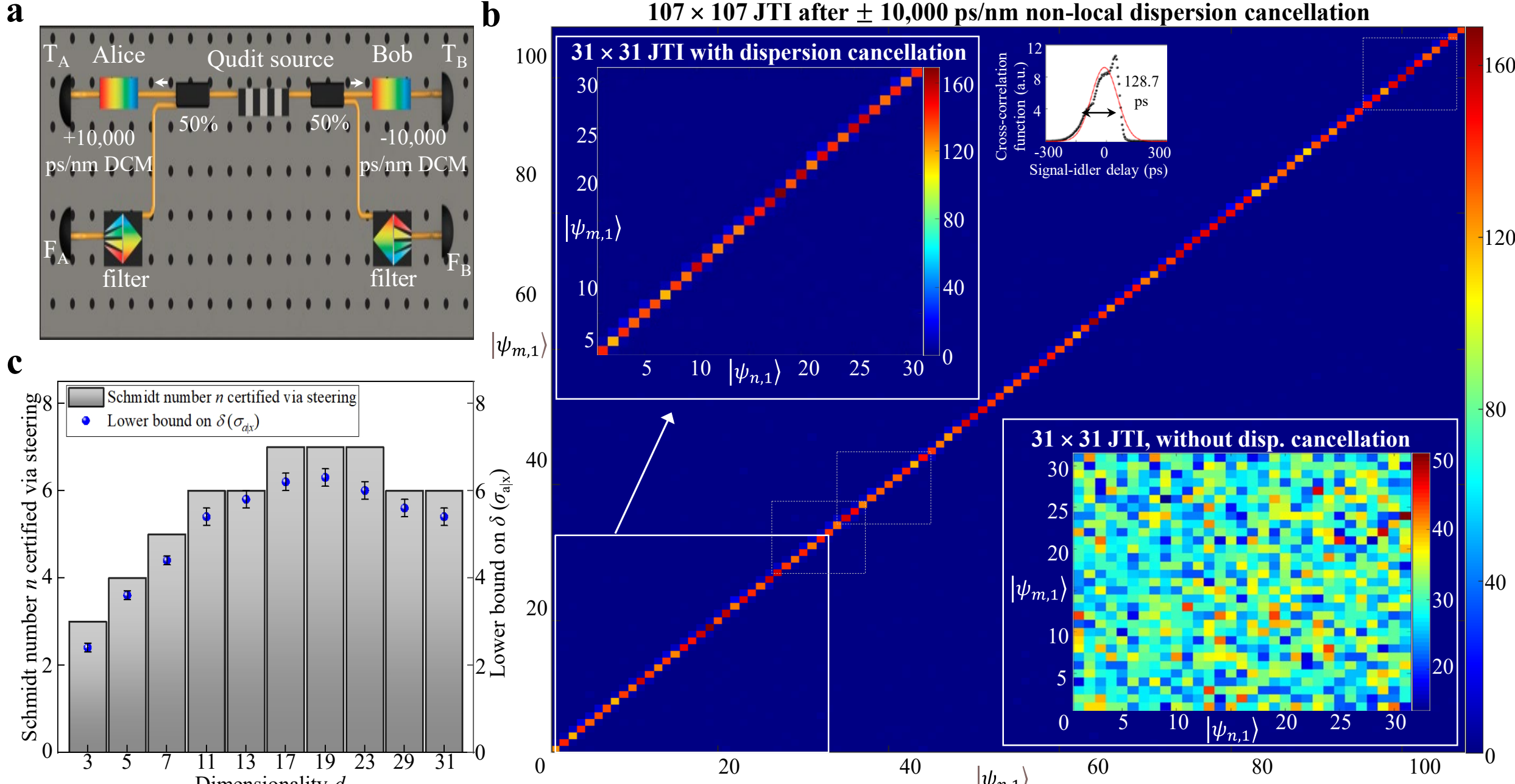

**Figure 3 | Time-frequency high-dimensional entanglement and quantum steering preservation after ± 10,000 ps/nm non-local dispersion cancellation. a,** Experimental setup for demonstrating time-frequency high-dimensional entanglement and quantum steering preserving after non-local dispersion cancellation. The ± 10,000 ps/nm dispersion emulator (compensator) perform non-local dispersion cancellation between $T_A$ and $T_B$ . Discretized JSI is measured using frequency filters. **b,** Here we show the two-photon coincidence counts from 107 × 107 dimensional discretized JTI after ± 10,000 ps/nm nonlocal dispersion cancellation. The bin-width $\tau$ is chosen at 600 ps to fully cover the entire two-photon correlation peak (top-middle inset), and the number of bins $N$ is 107, consistent with the temporal frame size in Figure 2b. After non-local dispersion cancellation, the FWHM of temporal correlation peak is observed to be ≈ 128.7 ps, and the asymmetry profile comes from the imperfect non-local dispersion cancellation. For a 107-dimensional JTI, we observe the coincidence counts of temporal correlated photons are consistent with the level of coincidence counts for a 31-dimensional JTI that we show in the top-left inset, supporting the dispersion cancellation approach. By using SNSPDs with lower timing jitter, we can use even smaller bin-widths to cover whole correlation-peak while continuing increasing number of bins $N$ to scale up the dimensionality. For comparison, bottom-right inset is a 31-dimensional JTI when the +10,000 ps/nm dispersion is not compensated. The duration of measured coincidence counting is 3 seconds. No subtraction of background or accidental counts is performed. Details on the after-dispersion cancellation JSI, high-dimensional entanglement witness using fidelity bound and entanglement-of-formation bound, and maximum certified entanglement dimensionality $d_{ent}$ are noted in the Supplementary Section IV (the transported maximum quantum state fidelity $\tilde{F}(\rho,\Phi)$, entanglement-of-formation $E_{oF}$, we observed are 93.1 ± 0.3%, and 2.5 ± 0.1 ebits, respectively). The uncertainty in fidelity is calculated from measurement of each data set, assuming Poisson statistics. **c,** We use our discretized 31 × 31 dimensional time-frequency bases to certify high-dimensional quantum steering after non-local dispersion cancellation. For a local dimension of 19, we extract SR's lower bound $\delta(\sigma_{a|x})$ of 6.3 ± 0.2. Hence, the certified Schmidt number $n$ is 7 that demonstrates a transported 7-dimensional quantum steering.

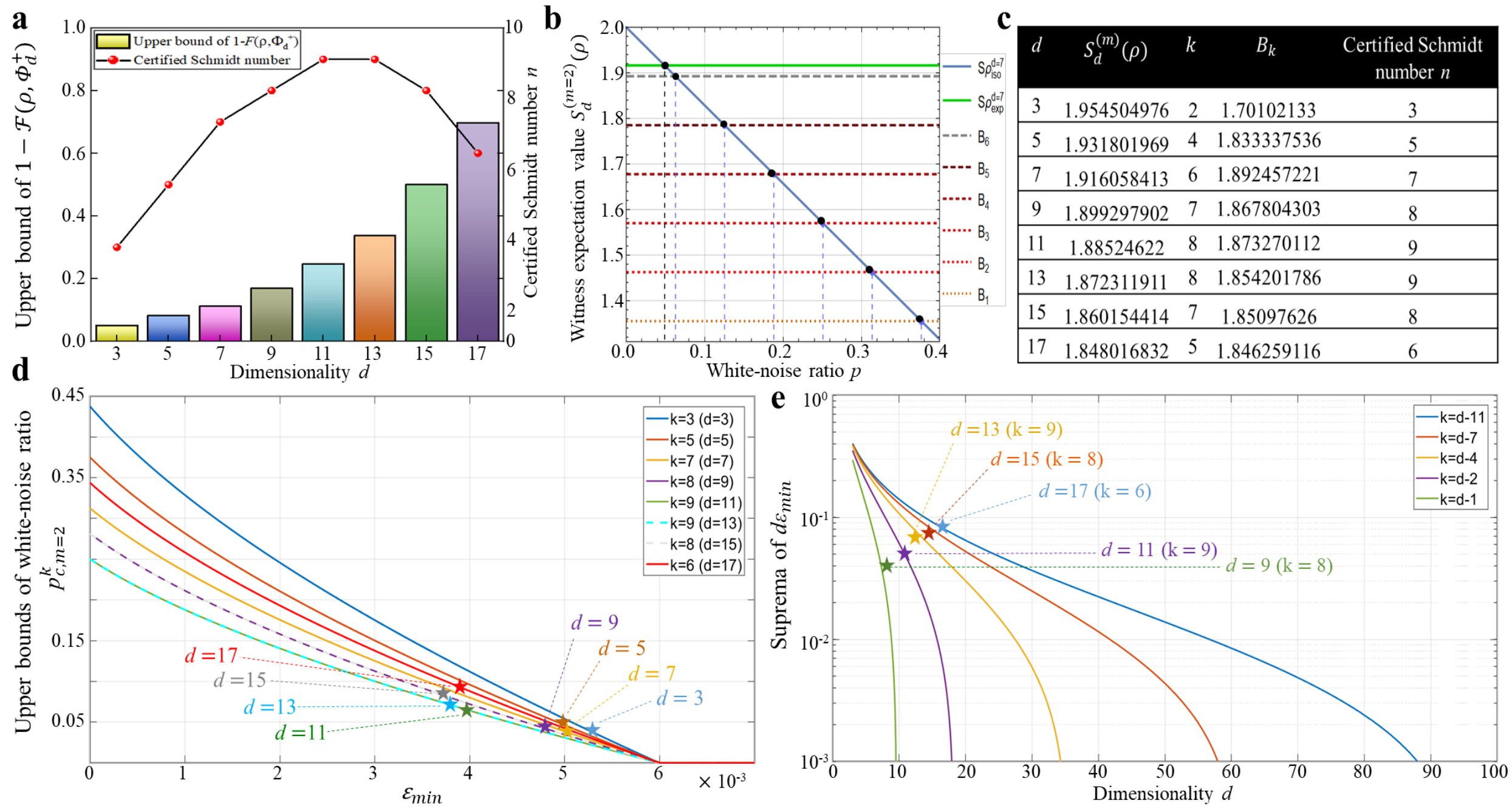

| $d$ | $S_d^{(m)}(\rho)$ | $k$ | $B_k$ | Certified Schmidt number $n$ |
|---|---|---|---|---|
| 3 | 1.954504976 | 2 | 1.70102133 | 3 |
| 5 | 1.931801969 | 4 | 1.833337536 | 5 |
| 7 | 1.916058413 | 6 | 1.892457221 | 7 |
| 9 | 1.899297902 | 7 | 1.867804303 | 8 |
| 11 | 1.88524622 | 8 | 1.873270112 | 9 |
| 13 | 1.872311911 | 8 | 1.854201786 | 9 |
| 15 | 1.860154414 | 7 | 1.85097626 | 8 |
| 17 | 1.848016832 | 5 | 1.846259116 | 6 |

**Figure 4 | High-dimensional entanglement certification via new witness for time-frequency bases. a,** The certified Schmidt numbers $n$ and upper bounds of $1-\mathcal{F}(\rho,\Phi_d^+)$, the infidelity with the maximally entangled state obtained using Schmidt-number witness from [83], are plotted against the local dimension $d$. Here we witness a Schmidt number of 9 at the qudit source for a local dimension $d$ of 13. **b,** We present the witness expectation value $S_d^{(m)}(\rho)$ versus the white-noise ratio $p$, where the solid black dot is the experimental data from our time-frequency bases, and $m$ is 2 here. Various thresholds $B_k$ are plotted, and the corresponding certified Schmidt number is witnessed by violating the thresholds $B_k$ with experimental witness expectation value $S_d^{(m)}(\rho)$ at different dimensions. $S_d^{(m)}(\rho_{AB}^{iso})$ represents the witness expectation value of isotropic state, a qudit Bell state mixed with a certain amount of white-noise. In our experiments, the white-noise comes from the detector and background photon noise. As the time-frequency dimension of the measurement bases increases, the experimental system has higher white-noise, hence, the witness expectation value $S_d^{(m)}(\rho)$ decreases. For all the other experimental data, we presented them in Figure 4c. **c,** Witness expectation value $S_d^{(m)}(\rho)$, the witness upper bounds $B_k$ corresponding to the largest $k$ such that $S_d^{(m)}(\rho) > B_k$, and the highest certified Schmidt number $n$ ($n = k + 1$) in each local dimension. This is computed from Figures 2b top-left inset and 2c, and the cross-basis measurements in Figure S7. **d,** Upper bounds of the white-noise ratio $p_{c,m=2}^k$ (by assuming the underlying state to be isotropic) for experimentally certifying Schmidt number $n$ versus the theoretically worst-case bases-bias parameter $\varepsilon_{min}$ in local dimensionality $d$ of 17. Here we superimpose our measurement data (stars) onto the theoretical model from Equation (5), for local dimension $d$ from 3 to 17. Note that some datapoints will be on the same curve since they have the same certified Schmidt number (for example, the certified Schmidt number 9 is shared between $d$ = 11 and 13), but with different upper bounds of the white-noise ratios $p_{c,m=2}^k$. The projected white-noise ratios of the state $p$ (plotted as stars) of given measurement data are obtained from the experimental witness expectation value $S_d^{(m)}(\rho)$ at different dimensions. **e,** Suprema of the dimension-rescaled bases bias $d\varepsilon_{min}$ versus the local dimensionality $d$ using our time-

frequency bases. First, from panel **d**, we extract the bases-bias parameter $\varepsilon_{min}$ for each dimensionality $d$ from our experiment and theory. Then, by using the experimentally certified Schmidt number $n$ in panels **a** and **c**, we compute the best fit between theoretical curves and our measurement dataset in panel **e**. Here we superimpose our measurement data (stars) onto the theoretical model for local dimension $d$ from 9 to 17. The white-noise ratio $p^{k}_{c,m=2}$ we used for panel **e** is 0.058, close to the white-noise ratio $p$ of 0.05 from our measurements. The close correspondence between measurement and the theory supports that time-frequency bases can be used to certify high-dimensional entanglement.